# Local-field distribution in resonant composites: Green's-function formalism

Ying Gu and K. W. Yu
*Department of Physics, The Chinese University of Hong Kong, Shatin, New Territories, Hong Kong, China*

Hong Sun
*Department of Physics and Institute of Condensed Matter Physics, Shanghai Jiao Tong University, Shanghai 200 030, China*

The effective response depends sensitively on composite microstructure due to large fluctuations in the local-electric field. For metallic clusters embedded in a dielectric host, the local-field distributions are extremely inhomogeneous in space around the metallic clusters due to quasistatic resonance, leading to a large enhancement in the effective linear and nonlinear responses. In this paper, we propose a general method for computing the electric field of metallic clusters near resonance via a perturbation formalism. We illustrate the method by simple examples. [S0163-1829(99)10619-2]

## I. INTRODUCTION

The optical properties of granular materials have attracted much interest. Many experiments were performed on composites of small metallic particles embedded in a dielectric host and a large infrared absorption was observed.[1–3] Recently, the optical nonlinearity of nanostructured composites has attracted much attention.[4] In particular, the metallic clusters exhibit strong nonlinear-optical response when they are structured on the nanometer scale, through the local-field and geometric-resonance effects, reflected in the spectral function.[5–7] For such composites, it is known that the local-field distributions are extremely inhomogeneous in space around the metallic clusters.[6]

There exist very efficient numerical methods for computing the effective conductivity of composite materials,[5] but these methods do not allow calculations of the field distribution. To avoid numerical difficulty, the effective-medium theory[8] (EMT) was extended to the nonlinear response of percolating composites and fractal clusters.[9–11] For linear problem, EMT usually captures the essential physics. For optical nonlinearity, however, EMT has disadvantages typical for all mean-field theories, namely, it diminishes the local-field fluctuations and yields results that may only be regarded as lower bound of the accurate results.

In this paper, we compute the local-electric field near resonance via the Green's-function formalism.[7] The Kirchhoff equations will be recast as a systematic perturbation expansion in the cluster quantities. Our aim is to find the electrostatic Green's function of the clusters subject to a point source. The organization of the paper is as follows. In the next section, we formulate a perturbation expansion for the lattice Green's function. In Sec. III, we obtain the solution in the subspace associated with the set of clusters. In Sec. IV, we compute the electric field near a geometric resonance of the clusters. We then illustrate the general method by various simple examples in Sec. V. Finally, we discuss the implication of our results on an efficient numerical computation of electric fields on random impedance networks.

## II. PERTURBATION EXPANSION FOR LATTICE GREEN'S FUNCTION

Consider a binary network in which impurity bonds of admittance $\epsilon_1$ are employed to replace the bonds in an otherwise homogeneous network of identical admittance $\epsilon_2$. The admittance of each bond is generally complex and frequency dependent. In what follows, a cluster is defined as a finite connected set of impurity bonds on the lattice (see Fig. 1). We will consider a set of clusters on a two-dimensional square lattice. Our objective is to find the electrostatic Green's function (i.e., impulse response) of the cluster subject to a point source. The Green's function, if found, can be used to compute the general response of the clusters by the principle of superposition. For instance, the results for a uniform field can be reproduced by placing a positive source and a negative source concomitantly on both sides of the cluster and in the limit of an infinite separation between the sources.

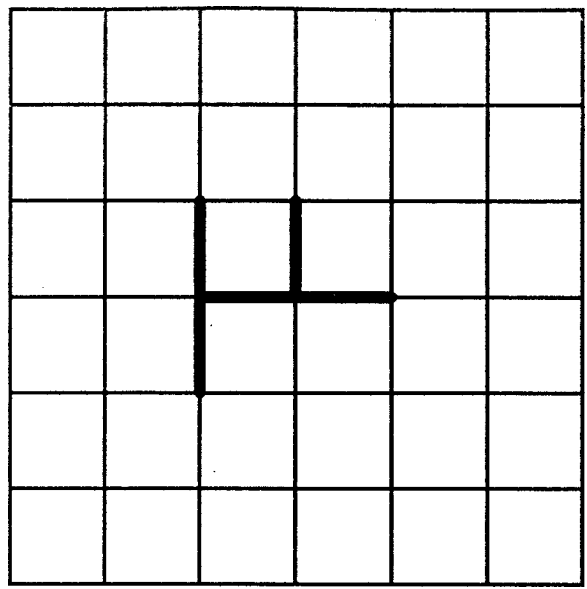

FIG. 1. Schematic diagram of a cluster (shown in thick lines) embedded in an infinite square network.

Suppose a unit point source is placed at site $\mathbf{0}=(0,0)$ outside a set of cluster of $n_s$ sites. The electric potential satisfies the Kirchhoff equation

$$\sum_{\mathbf{y(x)}} \epsilon_{\mathbf{x,y}}(F_{\mathbf{x,0}}-F_{\mathbf{y,0}})=\delta_{\mathbf{x,0}}, \qquad (1)$$

where $\mathbf{x}=(x_1,x_2)$ and $\mathbf{y(x)}$ denotes the four nearest-neighboring sites of $\mathbf{x}$; $F_{\mathbf{x,0}}$ is the electrostatic Green's function at $\mathbf{x}$ due to the point source at $\mathbf{0}$ and $\epsilon_{\mathbf{x,y}}=\epsilon_{\mathbf{y,x}}$ is the admittance of the bond joining the neighboring sites $\mathbf{x}$ and $\mathbf{y}$. Let $v=1-\epsilon_1/\epsilon_2$, Eq. (1) can be recast as

$$-\Delta F_{\mathbf{x,0}}=v\sum_{\mathbf{y}\in C(\mathbf{x})}(F_{\mathbf{x,0}}-F_{\mathbf{y,0}})+\frac{\delta_{\mathbf{x,0}}}{\epsilon_2}, \qquad (2)$$

where the notation $\mathbf{y}\in C(\mathbf{x})$ means that the bond $(\mathbf{x,y})$ belongs to the set $C$ of clusters, and $\Delta$ denotes the (finite difference) Laplace operator defined on the square lattice as

$$-\Delta Q_{\mathbf{x}}\equiv\sum_{\mathbf{y(x)}}(Q_{\mathbf{x}}-Q_{\mathbf{y}}), \qquad (3)$$

for any arbitrary physical quantity $Q_{\mathbf{x}}$. Equation (2) admits a formal solution

$$F_{\mathbf{x,0}}=\frac{G_{\mathbf{x,0}}}{\epsilon_2}+v\sum_{\mathbf{y}\in C}\sum_{\mathbf{z}\in C(\mathbf{y})}G_{\mathbf{x,y}}(F_{\mathbf{y,0}}-F_{\mathbf{z,0}}), \qquad (4)$$

written in terms of the Green's function $G_{\mathbf{x,y}}$ of the Laplace operator on the infinite square lattice, i.e.,

$$-\Delta G_{\mathbf{x,y}}=\delta_{\mathbf{x,y}} \text{ with } G_{\mathbf{x,x}}=0. \qquad (5)$$

The salient properties of $G_{\mathbf{x,y}}$ were reviewed in the appendix of Ref. 7. Equation (4) can be simplified by defining a matrix M

$$M_{\mathbf{x,y}}=\sum_{\mathbf{z}\in C(\mathbf{y})}(G_{\mathbf{x,y}}-G_{\mathbf{x,z}}), \qquad (6)$$

so that

$$F_{\mathbf{x,0}}=\frac{G_{\mathbf{x,0}}}{\epsilon_2}+v\sum_{\mathbf{y}\in C}M_{\mathbf{x,y}}F_{\mathbf{y,0}}. \qquad (7)$$

The first term on the right-hand side is the unperturbed Green's function of a point source at site $\mathbf{0}$ in the absence of the clusters, while the second term describes the perturbation due to the clusters. It should be remarked that M is generally an $\infty\times n_s$ matrix for an infinite network while $\mathbf{F}$ and $\mathbf{G}$ are column vectors of the Green's functions. We will look for solutions of Eq. (7) both for $\mathbf{x}\in C$ and $\mathbf{x}\notin C$.

### III. SOLUTION IN THE CLUSTER SUBSPACE

For the case $\mathbf{x}\in C$, it is more convenient to examine the subspace associated with the set of clusters. Let $\tilde{M}$, $\tilde{\mathbf{F}}$, and $\tilde{\mathbf{G}}$ denote the respective quantities in the subspace, i.e., $\tilde{M}$ is an $n_s\times n_s$ submatrix of M, while $\tilde{\mathbf{F}}$ and $\tilde{\mathbf{G}}$ are restricted to the $n_s$ cluster sites. For $\mathbf{x}\in C$, Eq. (7) becomes

$$\tilde{F}_{\mathbf{x,0}}=\frac{\tilde{G}_{\mathbf{x,0}}}{\epsilon_2}+v\sum_{\mathbf{y}\in C}\tilde{M}_{\mathbf{x,y}}\tilde{F}_{\mathbf{y,0}}, \qquad (8)$$

that can be rewritten as

$$\sum_{\mathbf{y}\in C}[\tilde{\delta}_{\mathbf{x,y}}-v\tilde{M}_{\mathbf{x,y}}]\tilde{F}_{\mathbf{y,0}}=\frac{\tilde{G}_{\mathbf{x,0}}}{\epsilon_2}, \qquad (9)$$

which can readily be inverted to yield $\tilde{\mathbf{F}}$, by solving a set of $n_s$ linear simultaneous equations. Substituting the solution of Eq. (9) into Eq. (7), the Green's function of all sites in the lattice can be obtained. The geometric resonance is characterized by a nontrivial solution of Eq. (9) even in the absence of an external field. In which case, $\tilde{\mathbf{F}}$ is dominated by one of the normal modes. The electric field around the cluster can be expressed in terms of the eigenvectors of the normal mode. To this end, let $s=1/v=\epsilon_2/(\epsilon_2-\epsilon_1)$, the $n_R$ real eigenvalues of $\tilde{M}$ lie in the range $0\leq s\leq 1$.[5] Only those nontrivial eigenvalues ($s\neq 0$, or 1) correspond to real physical resonances. For a particular eigenvalue $s_m$ of $\tilde{M}$, the corresponding right and left eigenvectors are denoted by $\tilde{\mathbf{R}}_m$ and $\tilde{\mathbf{L}}_m$ with components $\tilde{R}_{m,\mathbf{x}}$ and $\tilde{L}_{m,\mathbf{x}}$ in the representation of Eq. (6), respectively. These eigenvectors form a complete set and obey the orthonormal relation

$$\tilde{\mathbf{L}}_m\cdot\tilde{\mathbf{R}}_n=\sum_{\mathbf{x}\in C}\tilde{L}_{m,\mathbf{x}}\tilde{R}_{n,\mathbf{x}}=\delta_{m,n}, \quad (m,n=1,\ldots,n_R). \qquad (10)$$

There is always a trivial eigenvalue $s=0$ associated with the submatrix $\tilde{M}$; the corresponding (unnormalized) right eigenvector is the column vector $(1,1,\ldots,1)^T$. Hence, we obtain an identity by virtue of Eq. (10)

$$\sum_{\mathbf{x}\in C}\tilde{L}_{m,\mathbf{x}}=0. \qquad (11)$$

### IV. LOCAL-FIELD DISTRIBUTION NEAR RESONANCE

We are now in a position to solve for the Green's function near resonance. Let us write $\tilde{\mathbf{F}}$ as a linear combination of the eigenvectors $\tilde{\mathbf{R}}$'s

$$\tilde{\mathbf{F}}=\sum_{n=1}^{n_R}A_n(s)\tilde{\mathbf{R}}_n. \qquad (12)$$

Near resonance, $s$ is close to one of the eigenvalue $s_m$ of the submatrix $\tilde{M}$. Multiplying Eq. (8) by the left eigenvector $\tilde{\mathbf{L}}_m$, we obtain

$$A_m(s)=\frac{s}{\epsilon_2(s-s_m)}\tilde{\mathbf{L}}_m\cdot\tilde{\mathbf{G}}. \qquad (13)$$

Hence, for $\mathbf{x}\in C$,

$$\tilde{\mathbf{F}}=\sum_{n=1}^{n_R}\frac{s}{\epsilon_2(s-s_n)}\left(\sum_{\mathbf{y}\in C}\tilde{L}_{n,\mathbf{y}}\tilde{G}_{\mathbf{y,0}}\right)\tilde{\mathbf{R}}_n. \qquad (14)$$

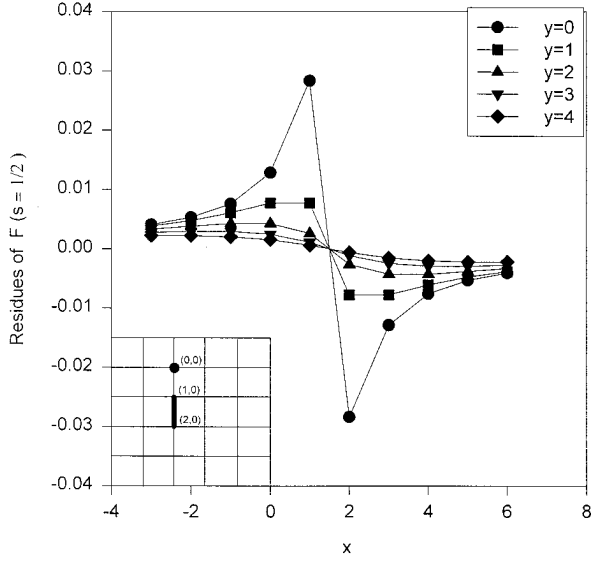

FIG. 2. The residue of the Green's function around a one-bond cluster plotted against the position $x$ along the bond for several different values of $y$ subject to a point source at $s=\frac{1}{2}$. The lines are guides to the eyes. In the inset, the cluster and the point source are shown.

When $s$ is very close to $s_m$, the local-field magnitude diverges, leading to an enhanced optical nonlinearity. We may define the residue of $\widetilde{F}_{\mathbf{x},\mathbf{0}}$ as

$$\text{Residue }(\widetilde{F}_{\mathbf{x},\mathbf{0}}) = \lim_{s \to s_m}(s-s_m)\widetilde{F}_{\mathbf{x},\mathbf{0}} = \frac{s_m}{\epsilon_2}\widetilde{R}_{m,\mathbf{x}}\left(\sum_{\mathbf{y}\in C}\widetilde{L}_{n,\mathbf{y}}\widetilde{G}_{\mathbf{y},\mathbf{0}}\right). \quad (15)$$

For $\mathbf{x} \notin C$, however, we obtain $\mathbf{F}$ from Eq. (7)

$$F_{\mathbf{x},\mathbf{0}} = \frac{G_{\mathbf{x},\mathbf{0}}}{\epsilon_2} + \frac{1}{\epsilon_2}\sum_{n=1}^{n_R}\frac{1}{s-s_n}\left(\sum_{\mathbf{y}\in C}\widetilde{L}_{n,\mathbf{y}}\widetilde{G}_{\mathbf{y},\mathbf{0}}\right)\left(\sum_{\mathbf{z}\in C}M_{\mathbf{x},\mathbf{z}}\widetilde{R}_{n,\mathbf{z}}\right). \quad (16)$$

It should be remarked that Eqs. (14) and (16) describe a generalized image problem in electrostatics. When the point of observation $\mathbf{x}$ is outside the cluster, there are two contributions to the Green's function. The first one arises from the point source as described by the first term of Eq. (16), whereas the second term describes the image contribution arising from the polarization of the clusters. When the point of observation is inside the cluster, however, there is only one term, i.e., the contribution from the screened source. Again, when $s$ is very close to $s_m$, we obtain the residue of $F_{\mathbf{x},\mathbf{0}}$

$$\text{Residue }(F_{\mathbf{x},\mathbf{0}}) = \frac{1}{\epsilon_2}\left(\sum_{\mathbf{y}\in C}\widetilde{L}_{m,\mathbf{y}}\widetilde{G}_{\mathbf{y},\mathbf{0}}\right)\left(\sum_{\mathbf{z}\in C}M_{\mathbf{x},\mathbf{z}}\widetilde{R}_{m,\mathbf{z}}\right). \quad (17)$$

Hence, the residue of $\mathbf{F}$ can be expressed as a separate product of a sum of the cluster property (i.e., independent of the point of observation $\mathbf{x}$) and another sum that depends on the point of observation.

The results for a uniform field can be reproduced by placing a positive source and a negative source concomitantly on both sides of the cluster. In the limit of an infinite separation between the sources, we obtain the following result[7]

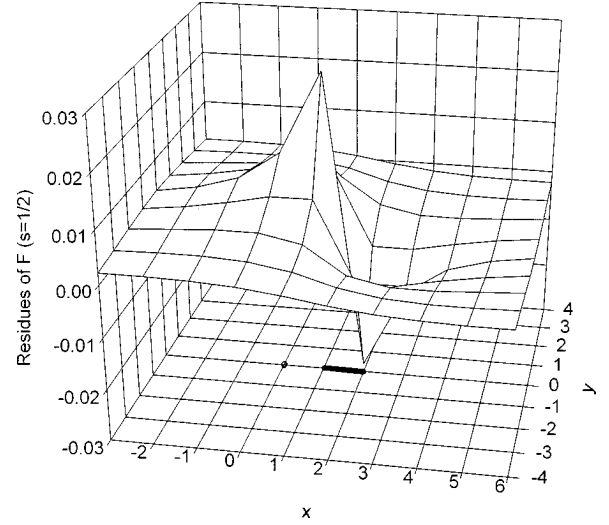

FIG. 3. Same as Fig. 2, but in a three-dimensional plot. The bond is shown as a thick line while the source is located at (0,0).

$$A_m(s) \to \frac{sE}{s-s_m}\sum_{\mathbf{x}\in C}x_1\widetilde{L}_{m,\mathbf{x}}, \quad (18)$$

where $E$ is the magnitude of the uniform applied field with $x_1$ being the coordinate of the site $\mathbf{x}$ along the applied field.

## V. EXAMPLES OF SIMPLE CLUSTERS

In this section, as an illustration of the general formulation, we compute the Green's function $\mathbf{F}$ for various simple clusters. We will study a one-bond cluster and various two-bond clusters because recent numerical simulation results on random impedance networks[12] showed that in the limit of a small volume fraction of metallic bonds, the optical-absorption spectrum is dominated by isolated clusters of a few bonds (or lattice animals).

### A. One-bond cluster

For a single bond [type (a) cluster as coined in Ref. 12] placed from site (1,0) to site (2,0) as shown in Fig. 2 with the source being placed at site (0,0), the submatrix reads

$$\widetilde{M} = \begin{pmatrix} G_0-G_1 & G_1-G_0 \\ G_1-G_0 & G_0-G_1 \end{pmatrix}, \quad (19)$$

where $G_0=0$ and $G_1=-\frac{1}{4}$. Since we always place the source at site $\mathbf{0}=(0,0)$, we have omitted the subscript $\mathbf{0}$ associated with the Green's functions for simplicity of notation. Apart from the trivial eigenvalue $s=0$, there is a nontrivial eigenvalue $s=\frac{1}{2}$, the normalized right and left eigenvectors associated with which are

$$\widetilde{R}_1 = \left(\frac{1}{\sqrt{2}}, -\frac{1}{\sqrt{2}}\right)^T, \quad \widetilde{L}_1 = \left(\frac{1}{\sqrt{2}}, -\frac{1}{\sqrt{2}}\right). \quad (20)$$

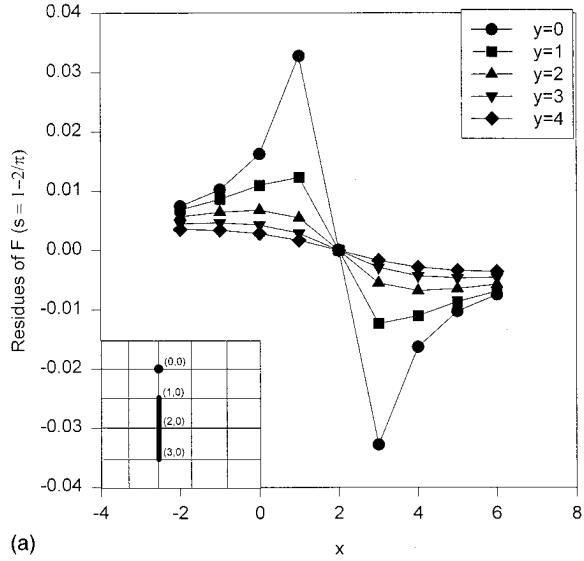

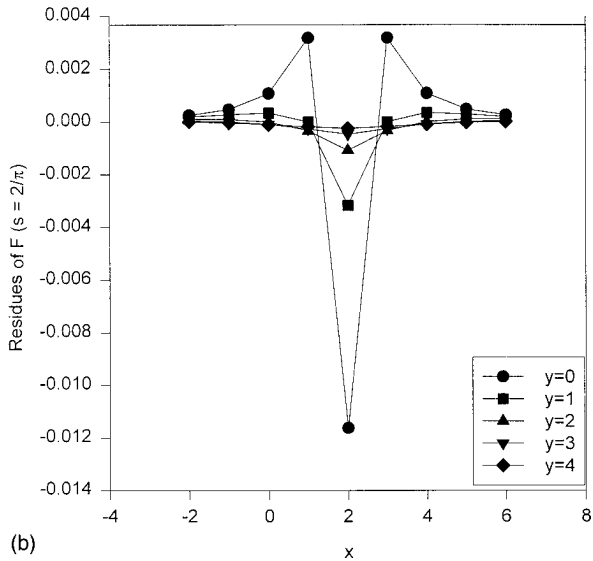

FIG. 4. The residue of the Green's function around a linear cluster plotted against the position $x$ along the bonds for several different values of $y$ subject to a point source at (a) $s = 1 - 2/\pi$ and (b) $s = 2/\pi$. Note the different behavior between the two cases. The lines are guides to the eyes. In the inset, the cluster and the point source are shown.

One can check that the identity [Eq. (11)] is obeyed. Without loss of generality, we let $\epsilon_2 = 1$. When $x \in C$, the residue of the Green's function of the one-bond cluster can be written as

$$\text{Residue } (\tilde{F}_x) = \lim_{s \to 1/2} (s - \tfrac{1}{2}) \tilde{F}_x = s_1 \tilde{R}_{1,x} (\tilde{L}_{1,1} G_1 + \tilde{L}_{1,2} G_2), \tag{21}$$

where $G_2 = 2/\pi - 1$. Hence, we have

$$\text{Residue } (\tilde{F}_1) = \frac{3}{16} - \frac{1}{2\pi}, \quad \text{Residue } (\tilde{F}_2) = -\left(\frac{3}{16} - \frac{1}{2\pi}\right). \tag{22}$$

When $x \notin C$, we obtain the residue of $F_x$,

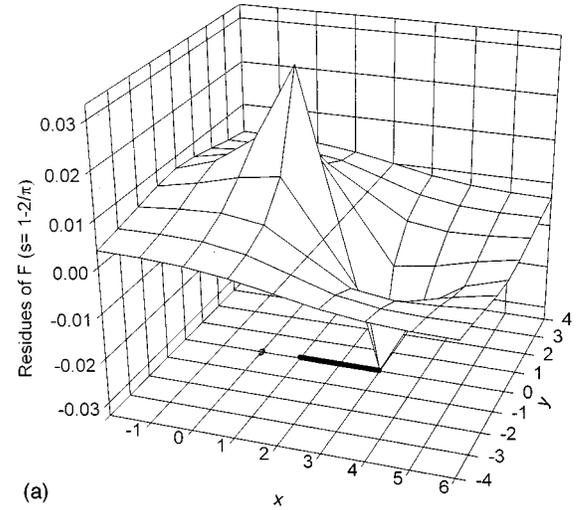

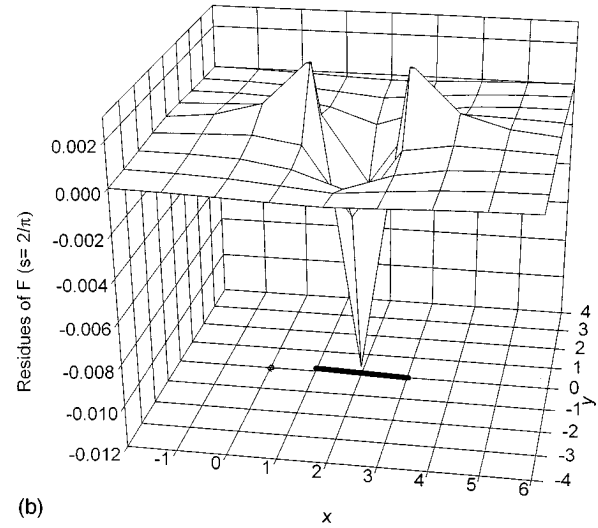

FIG. 5. Same as Fig. 4, but in a 3D plot. The linear cluster is shown as a thick line while the source is located at (0,0).

$$\text{Residue } (F_x) = (\tilde{L}_{1,1} G_1 + \tilde{L}_{1,2} G_2)(M_{x,1} \tilde{R}_{1,1} + M_{x,2} \tilde{R}_{1,2}). \tag{23}$$

Since the local-field distribution is symmetric about the bond, we only show the results of the Green's function on the right-hand side of the bond. As is evident from Fig. 2, there is a strong dipolar response along the bond at resonance. However, as we move farther away from the bond, the local field diminishes gradually, indicating that the resonance is strongly localized around the cluster.

Perhaps it is instructive to present the local-field distribution in a three-dimensional (3D) plot for better visualization. Hence, in Fig. 3, we plot the same results of the Green's function as a function of the position. We observe the same results, namely, there is a strong and localized dipolar response with a node at the mid point of the cluster.

### B. Various two-bond clusters

As yet another example, we consider a linear cluster from site (1,0) via site (2,0) to site (3,0) [type (c) cluster as coined in Ref. 12] as shown in Fig. 3(a); the point source is still placed at site (0,0). The associated submatrix reads

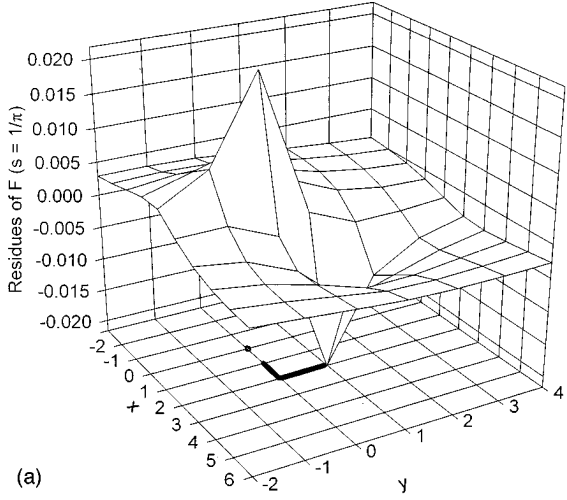

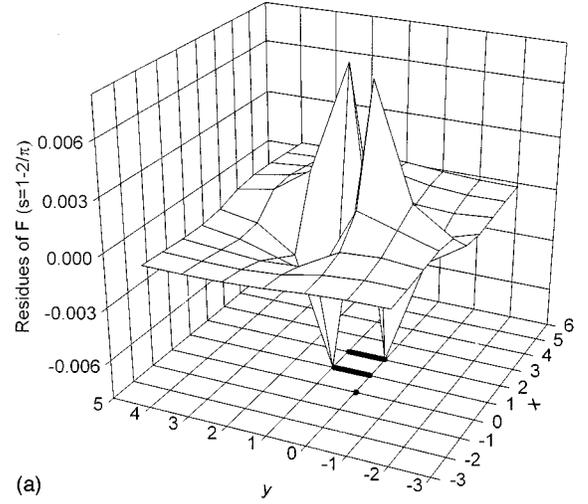

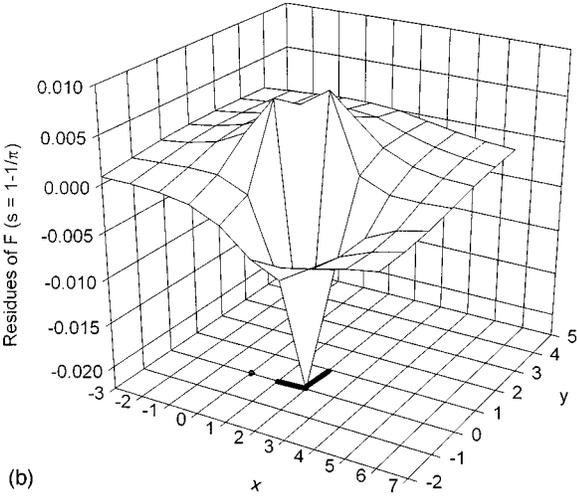

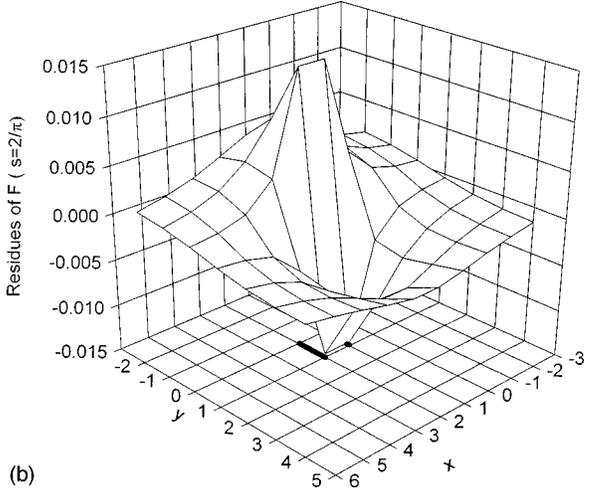

FIG. 6. The 3D plot of the residue of the Green's function around the type-(b) cluster against the position $x$ and $y$ subject to a point source at (a) $s=1/\pi$ and (b) $s=1-1/\pi$. The cluster is shown as a thick line while the source is located at (0,0).

FIG. 7. The 3D plot of the residue of the Green's function around the type-(d) cluster against the position $x$ and $y$ subject to a point source at (a) $s=1-2/\pi$ and (b) $s=2/\pi$. The cluster is shown as a thick line while the source is located at (0,0).

$$\tilde{M}=\begin{pmatrix} G_0-G_1 & 2G_1-G_0-G_2 & G_2-G_1 \\ G_1-G_0 & 2G_0-2G_1 & G_1-G_0 \\ G_2-G_1 & 2G_1-G_0-G_2 & G_0-G_1 \end{pmatrix}, \quad (24)$$

where $G_0=0$, $G_1=-\frac{1}{4}$, and $G_2=2/\pi-1$. There are two nontrivial eigenvalues at $s_1=1-2/\pi$ and $s_2=2/\pi$, respectively. For $s=1-2/\pi$, the right and left eigenvectors are

$$\tilde{R}_1=\left(\frac{1}{\sqrt{2}},0,-\frac{1}{\sqrt{2}}\right)^T, \quad \tilde{L}_1=\left(\frac{1}{\sqrt{2}},0,-\frac{1}{\sqrt{2}}\right). \quad (25)$$

We can readily check that Eq. (11) is obeyed. When $x \in C$,

$$\text{Residue } (\tilde{F}_x) = \lim_{s \to 1-\frac{2}{\pi}} \left[s-\left(1-\frac{2}{\pi}\right)\right] \tilde{F}_x$$

$$= \left(1-\frac{2}{\pi}\right) \tilde{R}_{1,x}(\tilde{L}_{1,1}G_1+\tilde{L}_{1,2}G_2+\tilde{L}_{1,3}G_3),$$

(26)

where $G_3=12/\pi-17/4$. Thus, we have

$$\text{Residue } (\tilde{F}_1) = -\text{Residue } (\tilde{F}_3)$$

$$= 2\left(1-\frac{2}{\pi}\right)\left(1-\frac{3}{\pi}\right), \quad \text{Residue } (\tilde{F}_2)=0.$$

(27)

When $x \notin C$, the residue of $F_x$ reads

$$\text{Residue } (F_x) = (\tilde{L}_{1,1}G_1+\tilde{L}_{1,2}G_2+\tilde{L}_{1,3}G_3)$$

$$\times (M_{x,1}\tilde{R}_{1,1}+M_{x,2}\tilde{R}_{1,2}+M_{x,2}\tilde{R}_{1,3}).$$

(28)

The residue of the Green's function at resonance is plotted in Fig. 3(a). The results are similar to those of the one-bond case, namely, there is a strong and localized dipolar response with a node at the middle site of the cluster.

While for $s=2/\pi$, the normalized right and left eigenvectors read

$$\tilde{R}_2 = (-0.25882, \ 0.94723, \ -0.25882)^T,$$

$$\tilde{L}_2 = (-0.41457, 0.82915, -0.41457), \quad (29)$$

where we have quoted the decimal approximation because of complicated analytic expressions. Again, we check that Eq. (11) is obeyed. When $x \in C$,

$$\text{Residue } (\tilde{F}_x) = \lim_{s \to \frac{2}{\pi}} \left( s - \frac{2}{\pi} \right) \tilde{F}_x$$

$$= \frac{2}{\pi} \tilde{R}_{2,x} (\tilde{L}_{2,1} G_1 + \tilde{L}_{2,2} G_2 + \tilde{L}_{2,3} G_3). \quad (30)$$

While for $x \notin C$,

$$\text{Residue } (F_x) = (\tilde{L}_{2,1} G_1 + \tilde{L}_{2,2} G_2 + \tilde{L}_{2,3} G_3)$$
$$\times (M_{x,1} \tilde{R}_{2,1} + M_{x,2} \tilde{R}_{2,2} + M_{x,2} \tilde{R}_{2,3}). \quad (31)$$

The residue of the Green's function is plotted in Fig. 4(b). As is evident from Fig. 4(b), there is also a strong and localized dipolar response along the bond. However, the response has some features quite different from that of Fig. 4(a), namely, there is an antinode at the middle site of the cluster.

Again, we present the local-field distribution in a 3D plot for better visualization. In Fig. 5, we plot the same results of the Green's function against position. For the other two-bond clusters, the computations of the Green's functions are essentially similar and the results are plotted for type-(b) cluster and type-(d) cluster in Figs. 6 and 7, respectively. Type-(b) cluster ranges from site (1,0) via site (2,0) to site (2,1) while type-(d) cluster is a set of two separate bonds with one bond ranging from site (1,0) to site (1,1) and the other bond from site (2,0) to site (2,1). Note that type-(b) cluster is self-dual and it has two nontrivial eigenvalues $s_1 = 1/\pi$ and $s_2 = 1 - 1/\pi$ while type-(d) cluster is the dual of type-(c) cluster and it has two nontrivial eigenvalues $s_1 = 1 - 2/\pi$ and $s_2 = 2/\pi$, identical with those of type-(c) cluster.

## DISCUSSION AND CONCLUSION

Here a few comments are in order regarding our Green's-function formalism. The present formalism deals with a perturbation expansion with respect to a set of cluster. As the cluster has fewer sites than the whole lattice, $\tilde{M}$ is a finite square matrix with size $n_s \times n_s$ being smaller than that ($N \times N$) of the full matrix for Kirchhoff equations. The Green's-function formalism thus provides an effective and practical means of computing the local-field distribution of clusters near a geometric resonance without the requirement of solving the Kirchhoff equation directly. In this connection, we note a similar formalism[13] in the bond representation, which however, did not deal with geometric resonance.

In summary, the conventional numerical methods are not applicable for computing the effective nonlinear response of composites because these methods do not allow calculations of the field distribution. In this work, the perturbation formalism allows us to compute the electric field near resonance via the Green's-function formalism.

## ACKNOWLEDGMENTS

This work was supported by the Research Grants Council of the Hong Kong SAR Government under Grant No. 4290/98P. K.W.Y. acknowledges useful conversations with Professor G.Q. Gu. H.S. acknowledges financial support by the Climbing (Pan-Deng) Program of the National Natural Science Foundation of China.

[1] D. B. Tanner, A. J. Sievers, and R. A. Buhrman, Phys. Rev. B **11**, 1330 (1975).

[2] C. G. Granqvist, R. A. Buhrman, J. Wyns, and A. J. Sievers, Phys. Rev. Lett. **37**, 625 (1976).

[3] R. P. Devaty and A. J. Sievers, Phys. Rev. Lett. **52**, 1344 (1984).

[4] See articles in *Nanostructured Materials: Clusters, Composites and Thin Films*, edited by V. M. Shalaev, ACS Symposium Series 679 (American Chemical Society, Washington, D.C., 1997).

[5] For a review, see D. J. Bergman and D. Stroud, in *Solid State Physics*, edited by H. Ehrenreich and D. Turnbull (Academic Press, New York, 1992), Vol. 146, p. 147.

[6] V. M. Shalaev, Phys. Rep. **272**, 61 (1996).

[7] For discrete network models, see J. P. Clerc, G. Giraud, J. M. Luck, and Th. Robin, J. Phys. A **29**, 4781 (1996).

[8] S. Kirkpatrick, Rev. Mod. Phys. **45**, 574 (1973).

[9] D. Stroud and P. M. Hui, Phys. Rev. B **37**, 8719 (1988).

[10] X. C. Zeng, P. M. Hui, D. J. Bergman, and D. Stroud, Phys. Rev. B **39**, 13 224 (1989).

[11] K. P. Yuen, M. F. Law, K. W. Yu, and P. Sheng, Phys. Rev. E **56**, R1322 (1997).

[12] M. F. Law, Y. Gu, and K. W. Yu, J. Phys.: Condens. Matter **10**, 9549 (1998).

[13] K. Wu and R. M. Bradley, Phys. Rev. E **49**, 1712 (1994).